\documentstyle[prd,aps,epsf,multicol]{revtex}
\begin{document}
\draft
%\preprint{Revised version, 22.3.97}

%%%%%%%%%%%%%%%%%%%%%%%%%%%%%%%%%%%%%%%%%%%%%%%%%%%%%%%%%%%%%%%%%%%%%%
\renewcommand{\narrowtext}{\begin{multicols}{2}}
\renewcommand{\widetext}{\end{multicols}}
\def\ba{\begin{eqnarray}}
\def\ea{\end{eqnarray}}

\def\ap{\approx}

%%%%%%%%%%%%%%%%%%%%%%%%%%%%%%%%%%%%%%%%%%%%%%%%%%%%%%%%%%%%%%%%%%%%%%%%%
\title{Axion cyclotron emissivity of magnetized white dwarfs 
       and neutron stars}

\author{ M. Kachelrie{\ss}, C. Wilke, and G. Wunner}

\address{Theoretische Physik I, Ruhr-Universit\"at Bochum, 
         D--44780  Bochum, Germany}

\maketitle

\begin{abstract}
The energy loss rate of a magnetized electron gas emitting axions $a$
due to the process $e^- \to e^- +a$ is derived for arbitrary magnetic 
field strength $B$. 
Requiring that for a strongly magnetized neutron star the axion 
luminosity is smaller than the neutrino luminosity we obtain
the bound $g_{ae} \lesssim  10^{-10}$ for the axion 
electron coupling constant. This limit is considerably weaker than the
bound derived earlier by Borisov and Grishina using the same 
method. Applying a similar argument to magnetic white dwarf stars
results in the more stringent bound 
$g_{ae}\lesssim 9 \cdot 10^{-13} 
(T/10^7 {\rm K})^{5/4} (B/10^{10} {\rm G})^{-2}$,
where $T$ is the internal temperature of the white dwarf. 
\end{abstract}

% axions, 'astroparticle physics', neutron stars, white dwarfs ect.

\narrowtext

%%%%%%%%%%%%%%%%%%%%%%%%%%%%%%%%%%%%%%%%%%%%%%%%%%%%%%%%%%%%%%%%%%%
\section{Introduction} 

The axion $a$ is a pseudoscalar boson introduced by Peccei and Quinn (PQ) 
to solve  the strong CP-problem in a natural way \cite{pe77}. It results
as a massless Goldstone boson from the spontaneous breaking of the 
PQ-symmetry by the VEV of a scalar field, $\langle\phi\rangle = f_{\rm PQ}$. 
As noted first by Weinberg and Wilczek \cite{ww}, the chiral
anomaly of QCD
induces, at low temperatures $T\lesssim\Lambda_{\rm QCD}\ap 200\:$MeV,
a mass $m_a \ap \Lambda_{\rm QCD}^2 / f_{\rm PQ}$ for the axion.
The numerical value of the axion mass is given by 
\begin{equation}
 m_a = \left( \frac{ 0.60\times 10^7 \mbox{GeV}}{f_{\rm PQ}} \right) 
 \mbox{eV} \:.
\end{equation}

Besides the coupling to two photons through the chiral anomaly, the
axion interacts with fermions by derivative couplings, which are all
inversely proportional to the axion mass. The interaction of axions
with electrons is described by 
\begin{equation}    \label{Lae}
 {\cal L}_{ae} = \frac{g_{ae}}{2m_e}\:
                 \bar\psi\gamma^\mu\gamma^5\psi \partial_\mu a  ,
\end{equation}
where the coupling constant 
\begin{equation}
 g_{ae}=\frac{m_e}{f_{\rm PQ}} \: c_e
\end{equation}
depends  on specific models through the effective PQ charge $c_e$. 
At tree level,
the electron has $c_e= 1/3 \cos\beta$  in the DSFZ axion model, 
while it has $c_e =0$ in the KSVZ model \cite{models}.

In invisible axion models, the axion mass $m_a$ is in principle
arbitrary, but in fact severely constrained by astrophysical and
cosmological considerations \cite{ko90,ra96}. 
The astrophysical bounds on $m_a$ arise because axion emission
is an additional energy loss mechanism for stars, and therefore
changes the stellar evolution.
Since the axion couplings to matter are proportional to
$m_a$, axion emission is suppressed for small  values of $m_a$.
Therefore astrophysics yields an upper bound for $m_a$. In
ref. \cite{ra95},  the bound
$m_a \lesssim 9 \cdot 10^{-3}$eV was derived  for a
DFSZ  axion by  studying the evolution of red giant stars.
On the other hand, cosmology provides a lower bound for $m_a$.
Axions would have been produced in the early Universe as coherent waves
or by the decay of axion strings. Therefore they 
could constitute cold dark matter. Requiring that axions do not
overclose the Universe yields the lower bound $ 10^{-5}$eV $\lesssim m_a$.

It is the purpose of the present work to examine the emission of
axions by electrons 
in the magnetic field $B$ of a strongly magnetized neutron star. 
The process considered, $e^- \to e^- + a$, is very similar to cyclotron
emission, $e^- \to e^- + \gamma$,  and, therefore, may be called axion
cyclotron emission. This process was first examined by Borisov and 
Grishina \cite{bo94} using a semiclassical expression for the
transition probability of the elementary process $e^- \to e^- + a$, 
valid a priori only in the limit of low magnetic field strengths,
$B\ll B_{\rm cr}= m_e^2 /e \ap 4.41 \times 10^{13}\:$Gauss, and of
ultrarelativistic electrons, $E\gg m_e$.
By comparing the axion cyclotron emission with neutrino cyclotron
emission $e^- \to e^- + \nu + \bar\nu$, they  
concluded that the axion electron coupling constant
is bounded from above by $g_{ae} \lesssim 5 \cdot 10^{-14}$. If this result, 
corresponding to $m_a \lesssim 6 \cdot 10^{-4}$eV for $c_e =1$, could be
confirmed or improved, the window in the axion mass range would become very
narrow, or even be closed. Consequently, it is imperative to recalculate
the axion cyclotron emissivity and to extend the results obtained in
ref. \cite{bo94} to arbitrary magnetic field strengths and electron energies. 
We derive the probability of the elementary
process $e^- \to e^- + a$ taking the magnetic field
exactly into account, and then numerically evaluate an
expression for the axion emissivity which is exact in the limit of
a completely degenerate Fermi-Dirac gas.
As the main result of this work we obtain the axion cyclotron luminosity 
of a degenerate star with arbitrary core magnetic field strength $B$. 
Although we confirm the analytical approximations of ref. \cite{bo94}
in the limit $B\ll B_{\rm cr}$ and $E\gg m_e$, the bound derived by us
for the axion electron coupling constant applying the energy-loss
argument to  neutron stars 
is three order of magnitudes weaker. The main reason for this is that 
all conventional neutrino emission mechanisms such as 
the modified URCA process, which we take into account, were neglected in 
ref. \cite{bo94}. % without justification.
Finally, we consider the energy loss of white dwarf stars due to axion
cyclotron emissivity. Requiring that for a white dwarf star the axion 
luminosity is smaller than the photon surface luminosity we obtain
the bound 
$g_{ae} \lesssim 9 \cdot 10^{-13} (T/10^7{\rm K})^{5/4} \,
(B/10^{10}{\rm G})^{-2}$.

%%%%%%%%%%%%%%%%%%%%%%%%%%%%%%%%%%%%%%%%%%%%%%%%%%%%%%%%%%%%%%%%%%%%%%%
\section{Elementary process}

We start by calculating the probability that an electron with polarization 
$\tau = \pm 1$ occupying an
excited Landau level $N$ emits an axion $a$ with momentum $k=(\omega, \vec k)$
and  undergoes a transition to the Landau level $N'$.
(Quantities describing the final electron are primed; compare also the
similar calculations of $e^- \to e^- + \gamma$ in ref. \cite{he82} 
for more technical details.)
The $S$-matrix element for this process,
\begin{eqnarray}
 S_{fi} \!  & = & \! \frac{ig_{ae}}{\sqrt{2\omega V}}  \int d^4 x \;
 \bar\psi^\prime (x)\gamma^{5} \psi (x) e^{ikx}  \,,
\end{eqnarray}  
follows directly from the Lagrangian   (\ref{Lae}) if we assume that there
exists only one Goldstone boson.

We choose the homogeneous magnetic field in $z$ direction, $\vec B= B
\vec e_z$, and as wave functions for the electron \cite{so68,he82}
\begin{equation} 
 \psi (x) = 
 \frac{(eB)^{1/4}}{\sqrt{L_y L_z}} \: 
 \frac{e^{-i(Et-p_y y-p_z z)}}{2\sqrt{E E_0 }}
  \:
 \left( \begin{array}{c}
 \Lambda_1\quad\phi_{N-1}(\xi ) \\
 \Lambda_2\quad \phi_{N}(\xi )  \\
 K_1\quad \phi_{N-1}(\xi ) \\ 
 K_2\quad \phi_{N}(\xi )  
\end{array}\right) . 
\end{equation}
Here, $\phi_N (\xi= x+ \frac{p_y}{eB})$ are the normalized eigenfunctions of 
the harmonic oscillator,  and we have introduced the abbreviations
\begin{eqnarray}
 \Lambda_1 &=&H\left( 1+\tau \right)+\Delta\left( 1-\tau \right)=\Pi_1^{\ast}
\\ 
\Lambda_2 &=&\Delta\left( 1+\tau \right)+H\left( 1-\tau \right)= \Pi_2^{\ast} 
\\
K_1&=& Z\left( 1+\tau\right) -\Gamma\left( 1-\tau\right) = M_1^{\ast} 
\\
K_2 &=& \Gamma\left( 1+\tau\right) -Z\left( 1-\tau \right) = M_2^{\ast} 
\end{eqnarray}
and
\begin{eqnarray}
H & = &\sqrt{(E+E_0 )(E_0 +m_e )}
\\ 
\Delta & = & -ip_z\sqrt{\frac{E_0 -m_e}{E+E_0}}
\\
Z & = & p_z\sqrt{\frac{E_0 +m_e}{E+E_0}}
\\
\Gamma & = & i\sqrt{(E+E_0)(E_0-m_e)}
\\
 E_0 & = & \sqrt{E^2 -p_z^2}=\sqrt{m_e^2 +2NeB} \:.
\end{eqnarray}

The time integration and two of the momentum integrations yield the usual 
conservation laws,
\begin{eqnarray}
 E & = & E^\prime + \omega \label{E} \\
 p_y & = & p_y^\prime + k_y \\
 p_z & = & p_z^\prime + k_z \label{p} \,,
\end{eqnarray}
while the $x$ integration results, up to a phase
factor, in Laguerre functions $I$,
\begin{equation}
 I_{N^\prime,N}(\kappa)= \sqrt{\frac{N^\prime !}{N !}} \;
  \kappa^{(N-N^\prime)/2} \, e^{-\kappa/2} \, 
  L_{N^\prime}^{N -N^\prime}(\kappa)
\end{equation}
for $N > N^\prime$. The argument of the $I$-functions is given by
$\kappa= (k\sin\theta)^2 / (2eB)$, $\theta$ is the
angle between $\vec B$ and $\vec k$, 
and the Laguerre polynomials $L_n^{n'} (x)$ are defined as in ref. \cite{ab}.
Using Eq. (\ref{E}) and (\ref{p}) we obtain the energy $\omega$ of the axion as
a function of $\theta$  
\ba
 \omega (\theta) & = & \Big\{ E-p_z \cos\theta -
                             \big[ \left( E-p_z \cos\theta \right)^2
\nonumber\\ &&
                          - 2 (N-N^\prime)eB \sin^2 \theta \big]^{1/2} 
                   \Big\} / \sin^2\theta \, ,
\ea
where, as in the following, we have neglected the axion mass $m_a$.%
\footnote{We will see below that $m_a \ll \omega \lesssim E_F$, where
$E_F = {\cal O}(m_e)$ is the Fermi energy of the degenerate electron gas.}

The decay width of an
electron in the Landau level $N$ into an Landau level $N^\prime$,  
\begin{equation}  \label{gamma} 
 d\Gamma^{N \to N'}_{\pm} (\vec k ) = 
  \lim_{T \to \infty }\, \frac{1}{T}
  \sum_{p^\prime_{y},p^\prime_{z},\tau' } \left| S_{\pm} \right|^2
  \frac{V}{(2\pi)^3}\; d^3 k   \; ,
\end{equation}
where $\pm$ denotes the (in general different)  decay widths of Landau states
with polarization $\tau =\pm$, becomes
\begin{equation}
 \Gamma^{N \to N' }_{\pm} = 
 \alpha_{ae} \sum_{\tau^\prime} \int_0^{\pi} d\theta \;
 \frac{\omega\sin\theta  \; \left| {\cal M} \right|^2  }
      {16^2 \:E E_{0} E^\prime_{0} \left( E' - p_z^\prime \cos\theta \right)}  
\end{equation}
with
\begin{eqnarray}   
{\cal M} & = &
  \Big( \Pi^\prime_{1} K_{1} - M^\prime_{1} \Lambda_{1} \Big) 
  I_{N^\prime -1 ,N -1} (\kappa) 
\nonumber\\ & + &
+  \Big( \Pi^\prime_{2} K_{2} - M^\prime_2 \Lambda_2 \Big) 
  I_{N^\prime ,N} (\kappa) 
\end{eqnarray}
and $\alpha_{ae}=g_{ae}^2/(4\pi)$.

%%%%%%%%%%%%%%%%%%%%%%%%%%%%%%%%%%%%%%%%%%%%%%%%%%%%%%%%%%%%%%%%%%%%%%%
\section{Axion emissivity}

The luminosity ${\cal L}_a$ emitted by ${\cal N}$ 
electrons occupying the volume
$V$ due to the process $e^- \to e^- + a$ is given by
\begin{equation}    \label{L}
 {\cal L}_a = 
     \lim_{T \to \infty }\, \frac{1}{T} \sum_{\lambda,\lambda^\prime} 
     \sum_{\vec k}   \omega |S_{\pm}|^2  \: {\cal S} \,,
\end{equation}
where the summation index $\lambda$ indicates the set of quantum numbers
$\lambda=\{ N,\tau, p_y, p_z \}$,
\begin{equation}
 {\cal S} = f (E)\left[ 1- f(E') \right] 
\end{equation}
and $f(E)$ are Fermi-Dirac distributions functions
\begin{equation}
 f (E) = \left[ \exp(\beta(E-\mu)) + 1\right]^{-1} \,.
\end{equation}
For the electron density $n_e = {\cal N}/ (L_x L_y L_z)$ and temperature
$T=\beta^{-1}$, the chemical potential $\mu$  is fixed through
\begin{equation}  \label{n_e}
 {\cal N} = \sum_\lambda f(E) =
         \frac{eB L_x L_y}{2\pi} \sum_{N=0}^\infty 
         \sum_\tau  % oder \left( 2-\delta_{N,0}\right)
         \int_{-\infty}^{\infty} d p_z \:\frac{L_z}{2\pi} \: f (E) \,.
\end{equation}
Here the factor $eB\, L_x L_y /(2\pi)$
takes into account the $p_y$ degeneracy of the Landau states.
For a nearly degenerate Fermi-Dirac gas, the $p_z$ integral in
Eq. (\ref{n_e}) can be
approximated (in a similar manner as in ref. \cite{la59}) by
\begin{eqnarray}
 n_e & = & \frac{eB}{2\pi^2} 
           \sum_{N=0}^{N_{\rm max}} \left( 2-\delta_{N,0}\right)
           \bigg[  \sqrt{\mu^2 -E_0^2}
\nonumber\\ & &
                - \frac{\pi^2}{6}\:\frac{E_0^2}{\sqrt{\mu^2 -E_0^2}^3}\: T^2 
                + {\cal O}\left(T^4 \right) 
       \bigg] \,,     
\end{eqnarray}
where $N_{\rm max} = {\rm int}\, [(\mu^2 -m_e^2 )/(2eB)] $.

The axion cyclotron emissivity $\varepsilon_a$ of electrons, 
i.e. the rate of energy loss per volume due to the process $e^- \to e^- + a$, 
follows from Eq. (\ref{L}) as
\begin{equation}    \label{eps,exakt}
 \varepsilon_a = \frac{eB}{(2\pi)^2}  \sum_{N=1}^\infty \sum_{N' < N} 
                 \sum_{\tau = \pm}
               \int_{-\infty}^\infty d p_z \int_0^\pi d\theta  
               \: \omega \frac{d\Gamma_{\pm}^{N \to N'} (\vec k)}{d\theta} 
               \: {\cal S} \,.
\end{equation}
The numerical evaluation of the terms  in Eq. (\ref{eps,exakt}) 
becomes  cumbersome already for moderate $N$. Therefore 
we take advantage of the degeneracy of the electron gas inside a
white dwarf or neutron star and employ an approximation commonly used in
calculations of neutrino emission rates: With the identity
\begin{equation}
 {\cal S} = \frac{1}{e^{\beta\omega}-1} \: \left[ f(E) -f(E^\prime) \right]
\end{equation}
and $f(E) \ap \theta (E-\mu)$, $f(E^\prime) \ap \theta (E^\prime-\mu)$,
the integration of a slowly varying function $g(E)$ of $E$ over $p_z$ 
becomes 
\begin{equation}
 \int_{-\infty}^\infty d p_z \: g(E) {\cal S} 
 \ap \frac{2 E_F}{p_z} \: g(E_{\rm F}) \: \frac{\omega}{e^{\beta\omega}-1} \,,
\end{equation}
and the Fermi energy $E_{\rm F}$ can be approximated by $E_{\rm F}\ap \mu$.
Now the summation over the discrete Landau levels $N$ breaks up at 
$N_{\rm max} = {\rm int} \, [(E_{\rm F}^2 -m_e^2)/(2eB)]$,
while the momentum $p_z$ of the initial electron is determined by 
$p_z = \sqrt{E_{\rm F}^2 - E_0^2}$. When $(E_{\rm F}^2-m_e^2)/(2eB)$
is an integer, the axion emissivity diverges because $p_z$ is zero or 
because, more physically speaking, the density of initial states is infinite. 
This can be remedied by taking into account the finite life time of
excited Landau states  (cf. ref. \cite{he82}) but is unimportant for
our purposes.

Although the two infinite sums over $N$ and $N'$ have been
replaced by
finite sums, for low magnetic field strengths or high densities it is
still necessary to compute Laguerre polynomials with high index. 
This can be avoided 
in the semiclassical, ultrarelativistic case ($B/B_{\rm cr} \ll 1,
E\gg m_e$) by the use of a Bessel function approximation for the
$I$-functions \cite{so68}, 
\begin{equation} \label{iasym}
I_{N^\prime ,N}\left( \kappa \right) =
   \frac{1}{\pi \sqrt{3}} \: \left( 1- \frac{\kappa}{\kappa_0} \right)^{1/2} 
     K_{1/3} (z)
\end{equation}
with
\begin{equation}
 \kappa_0 = \left( \sqrt{N} - \sqrt{N^\prime} \right)^2 
\end{equation}
and
\begin{equation}
 z=\frac{2}{3} \left(\kappa_0^2 \, N N^\prime \right)^{1/4} 
 \left( 1- \frac{\kappa}{\kappa_0} \right)^{3/2} \,.
\end{equation}
This asymptotic expansion is only valid for $\kappa\searrow \kappa_0$, 
i.e. $\theta \ap \pi/2$ is a necessary condition for its applicability. 
Therefore we evaluate $d\Gamma$ for $p_z =0$ and then carry out 
a Lorentz transformation to $p_z = \sqrt{E_{\rm F}^2 - E_0^2}$, viz.
\begin{equation}  \label{L1}
 \cos\theta_0 = \frac{E\cos\theta - p_z}{E - p_z \cos\theta}
\end{equation}
\begin{equation}
 d\Omega_0 = d\Omega \left( \frac{E_0}{E - p_z \cos\theta} \right)^2
\end{equation}
\begin{equation}      \label{L2}
 d\Gamma (k) = \frac{E_0}{E} \: d\Gamma_0 (k_0)  \,.
\end{equation}
Here quantities evaluated in the rest frame of the initial electron are  
denoted by the subscript $0$. Inserting the approximation for ${\cal S}$ into
Eq. (\ref{eps,exakt}) and using eq. (\ref{L1})--(\ref{L2}), we obtain
as final result for the axion emissivity in the limit of a degenerate
electron gas
\ba    \label{eps,approx}
 \varepsilon_a & = & \frac{eB}{2\pi^2} \:  
                 \sum_{N=1}^{N_{\rm max}}\sum_{N' < N}\sum_{\tau = \pm}
                 \int_0^\pi d\theta \sin\theta  \: \frac{E_0}{p_z}
\nonumber\\ &&
                 \left(\frac{E_0}{E - p_z \cos\theta} \right)^2
                 \frac{d\Gamma_{\pm 0}^{N \to N'}(\vec k)}{d\theta_0} \:
                 \frac{\omega^2}{e^{\beta\omega}-1} \,.     
\ea

Finally we examine the classical limit $p_z^2\ll m_e^2$ and $B\ll B_{\rm cr}$.
Inserting the approximations $E \ap m_e + NeB/m_e$, 
$E^\prime \ap m_e + N^\prime eB/m_e + p_z^{\prime\, 2}/ (2m_e)$,
and $\omega (\theta) \ap (N-N^\prime)eB/m_e$ in $|{\cal M}|^2$, we see
that $|{\cal M}|^2$ is of order ${\cal O}(B)$. Moreover, since 
$[e^{\beta\omega}- 1]^{-1} \ap T/\omega$, the axion emissivity is
proportional to $B^4 T$ in this limit. 
If the condition $N^2 B/B_{\rm cr} \ll 1$
is also valid, we can approximate the $I$-functions by
\begin{equation}   \label{I,nr}
 I_{N^\prime,N}(\kappa) =
  \kappa^{(N-N^\prime)/2}  \,
  \frac{ \sqrt{N!} }{ \sqrt{N^\prime !} (N-N^\prime)!} .
\end{equation}

%%%%%%%%%%%%%%%%%%%%%%%%%%%%%%%%%%%%%%%%%%%%%%%%%%%%%%%%%%%%%%%%%%
\section{Numerical results}

\subsection{Neutron stars}

\begin{center}
\begin{figure}
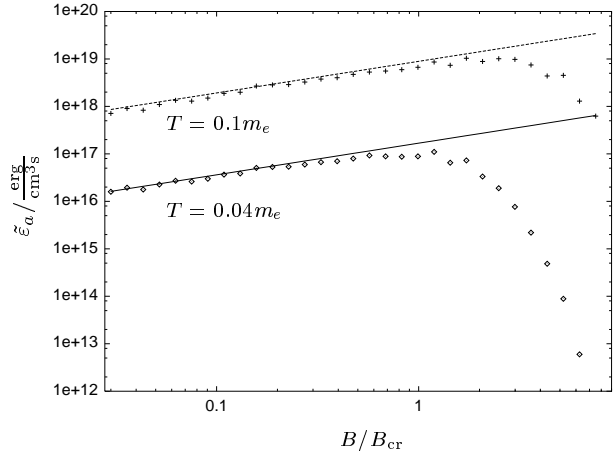

\bigskip
\caption{\label{ne32} 
  Axion emissivity $\tilde\varepsilon_a$ due to $e^- \to e^- + a$ 
  as a function of $B/B_{\rm cr}$ for $n_e = 10^{32}$cm$^{-3}$
  and for $T=0.04m_e$ (bottom) and $T=0.1m_e$ (top);
  lines: asymptotic formula, 
  crosses ($+$) and diamonds ($\diamond$): exact results).}
\end{figure}
\end{center}

\begin{center}
\begin{figure}
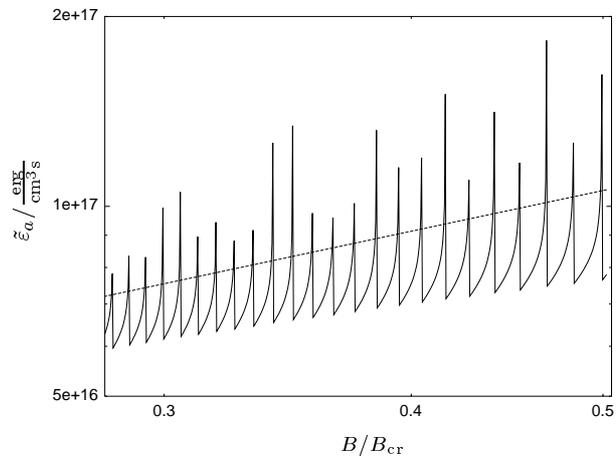

\bigskip
\caption{\label{detail} 
  Amplification of a small part 
  of Fig. \ref{ne32} showing the saw-tooth-like behavior of 
  the exact axion emissivity (for $T=0.04m_e$).}
\end{figure}
\end{center}

First, we will  establish that Eq. (\ref{eps,approx}) 
derived by us for general $B$ and $n_e$
% the axion emissivity $\varepsilon_a$ 
has the correct semiclassical limit  for $E\gg m_e$ and $B\ll B_{\rm cr}$
as given in ref. \cite{bo94}, 
\begin{equation}    \label{eps,cl}
 \varepsilon_a^{\rm cl}  = 8.67 \cdot 10^{-3} g_{ae}^2 m_e^5
     \left( \frac{p_F}{m_e} \right)^{-2/3} 
     \left( \frac{T}{m_e} \right)^{13/3} 
     \left( \frac{B}{B_{\rm cr}} \right)^{2/3} .
\end{equation}
Since for a relativistic degenerate Fermi gas and 
$1/3 \: E_F^2 \gg 2eB$ the usual relation 
$p_F = \sqrt[3]{3\pi^2 n_e}$ can be used \cite{ca68}, this approximation is
applicable for $n_e \gg m_e^3 / (3\pi^2) \ap 5.84 \cdot 10^{29}$cm$^{-3}$.
In Fig. \ref{ne32}, we compare our numerically accurate results with the 
approximate results for 
$\tilde\varepsilon_a = \varepsilon_a / \alpha_{1 {\rm eV}}$  as functions of 
$B/B_{\rm cr}$  for $n_e = 10^{32}$cm$^{-3}$, and for 
temperatures $T=0.04m_e$ and $T=0.1m_e$, respectively.  
(We use the emissivity scaled by 
$\alpha_{1{\rm eV}} = \alpha_{ae}/ (5.75 \cdot 10^{-22})$
to obtain a quantity independent of $m_a$.)
One recognizes that the exact results (crosses and diamonds)
reproduce the correct semiclassical  $B$ and $T$
dependence (straight lines) already for $B/B_{\rm cr}\lesssim 1$. 
To demonstrate the saw-tooth-like behavior caused by the root singularities
in the number density of the Landau states, an amplification of a small
excerpt of
Fig. \ref{ne32} is displayed in Fig. \ref{detail}. 
Note that these singularities are 
cancelled out on average in the semiclassical approximation.

\begin{center}
\begin{figure}
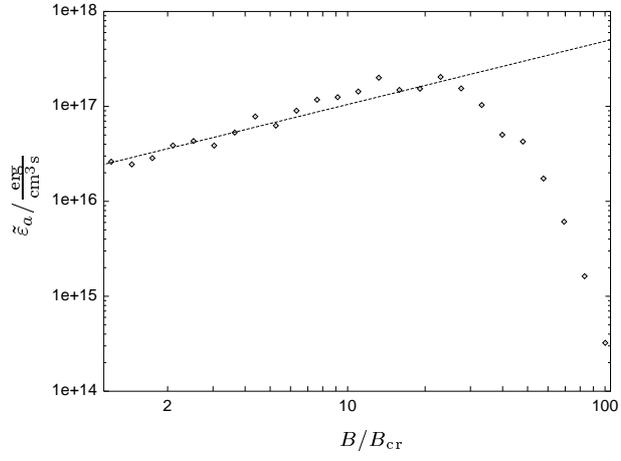

\bigskip
\caption{\label{rho0} 
  Axion emissivity $\tilde\varepsilon_a$  due to $e^- \to e^- + a$ 
  as a function of $B/B_{\rm cr}$ for 
  $n_e = 8.4 \cdot 10^{35}$cm$^{-3}$ and $T=0.04m_e$ 
  (dashed line: asymptotic formula, diamonds: exact results).}
\end{figure}
\end{center}

Fig. \ref{rho0} shows the magnetic field dependence of the 
axion emissivity $\tilde\varepsilon_a$ for   
$n_e = 8.4 \cdot 10^{35}$cm$^{-3}$ and $T=0.04m_e \ap 2.4 \cdot 10^8 $K. 
Assuming an ideal mixture of
nucleons and electrons in $\beta$-equilibrium, 
this value of the electron density $n_e$ corresponds to the
nuclear density $\rho_0 = 2.8 \cdot 10^{14}$g/cm$^3$. It can be
seen that the semiclassical approximation produces reasonable results even for 
$B/B_{\rm cr}\ap 20$, and not only for $B/B_{\rm cr}\ll 1$. 
A more suitable criterion for the applicability of Eq. (\ref{eps,cl})
is $N_{\rm max} \gg 1$, or, equivalently, for an ultrarelativistic
degenerate electron gas, % $E_F \gg m_e$,  
$E_F^2 \gg 2eB$. Except for a factor of three, this criterion
is the same as the condition derived in ref. \cite{ca68} for the
validity of the semiclassical treatment of a magnetized Fermi-Dirac gas.
For relatively low $N_{\rm max}$, the emissivity decreases
exponentially until for $2eB > \mu^2 - m_e^2$ all electrons populate
the Landau ground state (for $T=0$) and the emissivity becomes zero.

To obtain a bound for the coupling constant $g_{ae}$, we now compare
the axion luminosity with other processes important for the cooling
of neutron stars. 
A process specific to magnetized neutron stars is the cyclotron
emission of neutrino pairs by electrons, $e^- \to e^- + \nu + \bar\nu$.  
Kaminker {\it et al.} \cite{ka91}  obtained 
\begin{equation}   \label{nu,cyc}
 \varepsilon_{\nu,\rm cyc} = 
 (9 \times 10^{14} {\rm erg\;cm}^{-3}\; {\rm s}^{-1})
     \left( \frac{B}{10^{13}{\rm G}} \right)^{2} 
     \left( \frac{T}{10^9 {\rm K}} \right)^5 
\end{equation}
for the energy loss rate due to this process.
The bound $g_{ae} \lesssim 5\cdot 10^{-14}$ of ref. \cite{bo94} was derived
by requiring $\varepsilon_a^{\rm cl} \lesssim  \varepsilon_{\nu,\rm cyc}$  
under conditions appropriate for the outer crust of a neutron star.
Since only the total luminosity of stars is observable, a more
suitable -- but weaker -- criterion is to demand that the axion
luminosity ${\cal L}_a$ is smaller than the total neutrino luminosity $L_\nu$.
Other energy loss processes contributing to ${\cal L}_\nu$ are  
neutrino bremsstrahlung in the Coulomb field of ions \cite{fr79,sh83,st91},
\begin{equation}  \label{ion}
 \varepsilon_{\rm ions} = (2.1\times 10^{20} {\rm erg\;cm}^{-3}\; {\rm s}^{-1})
                       \frac{Z^2}{A}\: \frac{\rho}{\rho_0} \:
                       \left( \frac{T}{10^9 {\rm K}} \right)^6 
\end{equation}
and the modified URCA process,
\ba   \label{URCA}
 \varepsilon_{\rm URCA} & = & 
                       (2.7\times 10^{21} {\rm erg\;cm}^{-3}\;{\rm s}^{-1})
                       \left( \frac{m_n^*}{m_n} \right)^3 \: 
                       \frac{m_p^*}{m_p}  
\nonumber\\ &\times &                       
                       \left( \frac{\rho}{\rho_0} \right)^{2/3} \:
                       \left( \frac{T}{10^9 {\rm K}} \right)^8  \,.
\ea
The value $Z^2 /A$ is in the range $1-10$ \cite{ne73}, and
$m_n^* ,m_p^*$ are  effective masses of the Landau theory. In the
following we use $Z^2 /A =5$ and $m_p/m_p^* = m_p/m_p^* =1$. 
Recently, Pethick and Thorsson \cite{pe94} 
recalculated the neutrino bremsstrahlung process. 
Instead of treating the interaction of the electrons with the ions
immersed in a lattice in first-order perturbation theory, 
they used Bloch wave functions for the electrons 
and found that bremsstrahlung is exponentially suppressed at low
temperatures, $T\lesssim m_e$.

As a model neutron star we choose
a neutron star with radius $R=10\,$km, constant density $\rho$ equal to
nuclear density $\rho_0$ and temperature $T=0.04m_e$. 
Since the neutron star observations by ROSAT indicate that the
standard cooling scenario without exotic matter 
describes the cooling curves \cite{be95}, 
we assume the absence of a pion condensate.
Young pulsars have magnetic dipole fields 
$10^{12}{\rm G} \lesssim B_{\rm dipole} \lesssim 3\cdot 10^{13}$G. Since the
magnetic field strength inside the neutron star is  generally assumed to 
be stronger than $B_{\rm dipole}$ (cf. e.g. \cite{th93}),
we choose $B = {\cal O} (1-10B_{\rm cr})$. 

First we consider  the case where the core of the
neutron star is not superfluid.
Then the total neutrino luminosity is
${\cal L}_\nu = 4\pi/3 \; R^3 ( \varepsilon_{\rm ion} 
                       + \varepsilon_{\rm URCA} ) 
\ap 1.3 - 8.9 \cdot 10^{35} {\rm erg /s}$.
Here the first number takes into account
the exponentially suppression found in ref. \cite{pe94} by a factor 
$\ap 0.02$ for $\varepsilon_{\rm ion}$, while 
the second number uses $\varepsilon_{\rm ion}$ given by 
eq. (\ref{ion}).
The luminosity due to cyclotron emission of $\nu$ pairs 
(eq. (\ref{nu,cyc})) could be neglected in both cases.
For $T=0.04m_e$, we read off the axion emissivity 
$\tilde\varepsilon_a  \ap 10^{17} {\rm erg \: cm}^{-3} {\rm s}^{-1}\:$ 
from Fig. \ref{rho0}, and obtain for the axion luminosity 
${\cal L}_a = 4 \cdot 10^{35} \,  \alpha_{1 {\rm eV}} \,{\rm erg/s}$.
Requiring ${\cal L}_a < {\cal L}_\nu$, we arrive at the bound $g_{ae} \lesssim
5\cdot 10^{-11} - 1\cdot 10^{-10}$.

Next we comment on the case that in the core of the neutron star
superfluid protons form a type II 
superconductor \cite{sa89}. Then the initially homogenous magnetic field 
$\bar B$ becomes 
confined to quantized flux tubes (Abrikosov fluxoids). Each fluxoid carries
an elementary magnetic flux quantum $\phi_0=\pi/e$ and has the field profile
\begin{equation}   \label{fluxoid}
 B(r) \ap \frac{\phi_0}{2\pi\lambda^2}\:K_0 \left(\frac{r}{\lambda}\right) \:,
\end{equation}
where $\lambda$ is the penetration depth of the magnetic field 
and $K_0$ is a modified Bessel 
function. The above formula for $B(r)$ is valid if the proton-proton 
correlation length is much smaller than the distance $r$ from the fluxoid 
axis. The fluxoids are separated by the mean distance
$d_F = [2\phi_0/(\sqrt{3}\bar B)]^{1/2}$.
We can approximate the mean axion emissivity $\langle\varepsilon_a\rangle$ 
by the average over the lattice of fluxoids with a quasi-homogenous 
magnetic field given by eq. (\ref{fluxoid}),
\begin{equation}   \label{quasi-uniform}
 \langle\varepsilon_a \rangle 
 \ap \frac{2}{d_F^2} \int_0^{d_F} dr \:  r \varepsilon_a(r) \,,
\end{equation}
if the coherent interaction length $l_{\rm int}$ is much smaller than the
characteristic length scale $\lambda$ on which the magnetic field varies. 
Typical values for $\lambda$ are $\lambda\ap 30\,{\rm fm} \ap 0.1 m_e^{-1}$,
while $l_{\rm int}$ is given by the uncertainty principle,
$l_{{\rm int},x,y} \gtrsim 1/\sqrt{eB}$ and by
$l_{{\rm int},z} \gtrsim 1/p_z$. Thus $l_{\rm int}$ has the same order of 
magnitude as $\lambda$ and the quasi-uniform approximation is not well
justified. Therefore the approximation (\ref{quasi-uniform}) should be seen
only as a crude estimate.
For initial magnetic field strengths $\bar B = B_{\rm cr}$ and
$\bar B = 10B_{\rm cr}$ we obtain 
$ \langle\varepsilon_a \rangle 
\ap 3 \cdot 10^{15} {\rm erg \: cm}^{-3} {\rm s}^{-1}\:$ 
and
$ \langle\varepsilon_a \rangle 
\ap 3 \cdot 10^{16} {\rm erg \: cm}^{-3} {\rm s}^{-1}\:$,
respectively. This means that axion cyclotron emissivity is only weakly
suppressed by the superfluidity of protons, while the
neutrino emissivities (\ref{ion}) and (\ref{URCA})
are suppressed by a factor of $\exp ( -2 \beta\Delta_n )$, where $\Delta_n$
is the energy gap at $T=0$. On the other hand, new energy loss mechanisms
such as neutrino emission by scattering of electrons on fluxoids could become
important \cite{ka97}. Because of the uncertainties in both the axion and  
the neutrino emissivities, we do not consider it appropriate
to try to derive a rigorous bound for $g_{ae}$ in the superfluid case.

\subsection{White dwarfs}

We now apply the same line of arguments to magnetic white dwarfs. The internal
temperature $T$ of
a white dwarf with mass $M$ is related to its surface photon
luminosity ${\cal L}_\gamma$ by \cite{ra96,me52}
\begin{equation}
 \frac{{\cal L}_\gamma}{M} =  3.3 \cdot 10^{-3} \: 
                        \frac{\rm{erg}}{{\rm g}\,{\rm s}} \: 
                        \left( \frac{T}{10^7{\rm K}} \right)^{7/2} \,.
\end{equation}
About $5\%$ of white dwarfs have surface magnetic fields in the range
$10^{6}-10^{9}$G. Although the internal magnetic fields strengths are
assumed to be in the range $10^8 -10^{10}$ G, they could approach even
$10^{12}$ G \cite{sh83}.
%Their internal magnetic field strengths are
%assumed to be substantially higher. 

The mean density of a white dwarf depends on its total mass and chemical
composition but nevertheless a typical mean density is
$10^6$g/cm$^{3}$ corresponding to $n_e \ap 3 \cdot 10^{29}$cm$^{-3}$. 
Unfortunately, for this value of $n_e$ none of the two
approximations for the $I$-functions is applicable.
Therefore, we compute $\varepsilon_a$ for densities low enough such that 
the use of the approximation Eq. (\ref{I,nr}) is possible.
In Fig. \ref{wd}, $\tilde\varepsilon_a$ is shown for $n_e =10^{26}$cm$^{-3}$
and the temperatures $T=10^6$K, $T=10^7$K and $T=10^8$K, respectively,
together with the fit function 
\begin{equation}  \label{fit}
 \tilde\varepsilon_a^{\rm fit} = 2.6 \:\frac{\rm{erg}}{{\rm cm}^3{\rm s}} \: 
 \left( \frac{B}{10^9 {\rm G}} \right)^4
 \left( \frac{T}{10^7 {\rm K}} \right) \, .
\end{equation}           
The agreement with the assumed $B^4 T$ behavior is good. 
We want to give a simple argument in order to decide if eq. (\ref{fit}) over- 
or underestimates the true emissivity because of the use of the 
low density $n_e =10^{26}$cm$^{-3}$:
The ultrarelativistic approximation eq. (\ref{eps,cl}) yields,  for the values
$B=10^9 {\rm G}$, $T=10^7 {\rm K}$ and $n_e = 3 \cdot 10^{29}$cm$^{-3}$,
a much higher emissivity than $\tilde\varepsilon_a^{\rm fit}$. 
Therefore the non-relativistic approximation  
eq. (\ref{fit}) can be used as a {\em lower\/} bound for the true emissivity, 
if one assumes a smooth
transition between the non-relativistic and the ultrarelativistic 
asymptotic behavior.
      
We now require that magnetized white dwarfs do not emit 
more energy in axions than in photons, i.e. that
${\cal L}_\gamma /M \lesssim \varepsilon_a / \rho$, and obtain
\begin{equation}  \label{wd,limit}
 g_{ae}\lesssim 9 \cdot 10^{-13} \left( \frac{T}{10^7 {\rm K}} \right)^{5/4} 
                          \left( \frac{B}{10^{10} {\rm G}} \right)^{-2} \,.
\end{equation}
Since this bound is quite sensitive to $B$ and the knowledge of the
internal magnetic field strengths of white dwarfs is poor, it is hard
to derive a precise bound for $g_{ae}$. However, since 
$\varepsilon_a \propto T$, compared to ${\cal L}_\gamma \propto T^{7/2}$, 
axion emission becomes more and more
important during the cooling history of magnetized white
dwarfs. Therefore the fraction of magnetized white dwarfs among all
white dwarfs should diminish drastically for low enough luminosities.
In Table I, we compare the fraction of strongly magnetized white
dwarfs with the fraction of all white dwarfs in three different
temperature bins \cite{ra96,sc95}.
The distribution of hot strongly magnetized white dwarfs adapted from
Ref. \cite{sc95} spans approximately three order of magnitudes in
surface dipole field strength, 
$B_{\rm dipole} = 3\cdot 10^6 - 10^9$ G, and  has a maximum at 
$B_{\rm dipole}\ap 3\cdot 10^7$ G. It consists of only 18 stars, so some
caution in interpreting the data is appropriate. Nevertheless, the
fraction of magnetized white dwarfs in the last temperature bin is
considerably diminished compared to the total white dwarf population.
One {\em possible\/} explanation for this could be the additional
energy loss of magnetized white dwarfs due to axion cyclotron emission.

%Naturally, this limit applies equally well to other light pseudoscalar
%particles which couple as in eq. (\ref{Lae}) to electrons.

%%%%%%%%%%%%%%%%%%%%%%%%%%%%%%%%%%%%%%%%%%%%%%%%%%%%%%%%%%%%%%%%%%%%%%%
\section{Summary}

\begin{center}
\begin{figure}
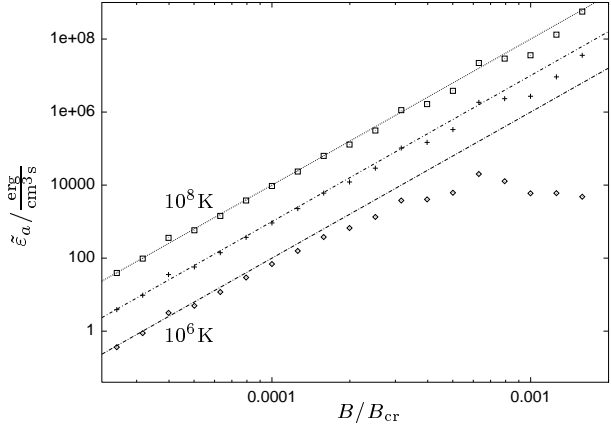

\bigskip
\caption{\label{wd} 
  Axion emissivity $\tilde\varepsilon_a$  due to $e^- \to e^- + a$ 
  as function of $B/B_{\rm cr}$ for 
  $n_e = 10^{26}$cm$^{-3}$ and $T=10^8$K, $T=10^7$K, $T=10^6$K   
  (lines: fit function Eq. (44), points: exact results).}
\end{figure}
\end{center}

We have derived the axion emissivity of a magnetized electron gas 
due to the process $e^- \to e^- +a$ for arbitrary magnetic 
field strengths $B$. 
Comparing the exact emissivity with the
semiclassical approximation obtained earlier, we have shown that this
approximation is not only valid for $B/B_{\rm cr} \ll 1$, but also under the
more general condition $E_F^2 \gg 2eB$. 
Applying axion cyclotron emission as an additional cooling 
mechanism to neutron stars and requiring that the axion luminosity is
smaller than the neutrino luminosity 
we could constrain the axion electron coupling constant to a value as 
small as $g_{ae} \lesssim {\cal O} (10^{-10})$. This bound is three orders of
magnitude  weaker than the
bound derived in ref. \cite{bo94} considering the same process.
This discrepancy  is not caused by differences in the calculation of
the axion emissivity but by different assumptions about neutron star
cooling. 
In the case of white dwarfs we could derive the more stringent limit
eq. (\ref{wd,limit}). 
We have noted that the lack of low-temperature magnetized
white dwarfs could be interpreted as signature of an additional
energy loss due to axion cyclotron emission.

%This corresponds to the mass limit
%$m_a \lsim  2 \:{\rm meV} \cos\beta 
%\left( \frac{T}{10^7 {\rm K}} \right)^{5/4} 
%\left( \frac{B}{10^{10} {\rm G}} \right)^{-2}$
%in the DSFZ model.

%%%%%%%%%%%%%%%%%%%%%%%%%%%%%%%%%%%%%%%%%%%%%%%%%%%%%%%%%%%%%%%%
\acknowledgments

During the initial phase of this work, MK was supported by grants from
Deutscher Akademischer Austauschdienst and Land Nordrhein-Westfalen.

%%%%%%%%%%%%%%%%%%%%%%%%%%%%%%%%%%%%%%%%%%%%%%%%%%%%%%%%%%%%%%%%%

\newpage

\begin{table}
\begin{tabular}{ccc}               
 $T_{\rm eff} [10^3 $K] &  
 Fraction of all WDs & Fraction of magnetized WDs \\
\hline
 40-80 & 1\% & 0\% \\
\hline
 20-40 & 23\% & 50\% \\
\hline
 12-20 & 76\% & 50\% \\
\end{tabular}
\bigskip
\caption{\label{tab}
Fraction of all white dwarfs (WDs) and of magnetized WDs in three
temperature bins.}  
\end{table}

\newpage
\widetext
%%% 1 inch =2.54 cm

\newpage

%\widetext


\begin{references}  
%
\bibitem{pe77} 
R. D. Peccei and H. R. Quinn, 
Phys. Rev. Lett. {\bf 38}, 1440 (1977); Phys. Rev. {\bf D16}, 1791 (1977).

\bibitem{ww} S. Weinberg, Phys. Rev. Lett. {\bf 40}, 223 (1978);
F. Wilczek, Phys. Rev. Lett. {\bf 40}, 271 (1978).

\bibitem{models} 
% DSFZ:
M. Dine, W. Fischler, and M. Srednicki, Phys. Lett. {\bf B104}, 199 (1981); 
A. P. Zhitnisky, Sov. J. Nucl. Phys. {\bf 31},260 (1980).
% KSVZ 
J.-E. Kim, Phys. Rev. Lett. {\bf 43}, 103 (1979);
M. A. Shifman, A. I. Vainshtein, and V. I. Zakharov, 
Nucl. Phys. {\bf B166}, 260 (1980).
%
For a review of different axion models and references see 
J.-E. Kim, Phys. Rep. {\bf 150}, 1 (1987).


\bibitem{ko90} 
%G. Raffelt, Phys. Rep. {\bf 198}, 1 (1990);
%For an up-date see G. Raffelt in 
%B. Guiderdoni, G. Greene, D. Hinds, J. Tran Thanh Van (eds.),
%{\it Proc. 15th Rencontres de Moriond\/} 
%(Edition Les Frontieres: 1995), 
%hep-ph/9502358.
E. W. Kolb and M. S. Turner, {\it The Early Universe\/}
(Addison-Wesley: Redwood City 1990).

\bibitem{ra96}
G. Raffelt, {\it Stars as Laboratories for Fundamental Physics\/}
(University of Chicago Press: Chicago 1996).

\bibitem{ra95} 
G. Raffelt and A. Weiss, Phys. Rev. {\bf D51}, 1495 (1995).

\bibitem{bo94} 
A. V. Borisov and V. Yu. Grishina, 
JETP {\bf 79}, 837  (1994).

\bibitem{so68} 
A. A. Sokolov and I. M. Ternov, 
{\it Synchrotron Radiation\/}     
(Akademie-Verlag: Berlin 1968). 

\bibitem{he82} 
H. Herold, H. Ruder, and G. Wunner, 
%Phys. Lett. {\bf 91B}, 272 (1982).
Astron. Astrophys. {\bf 115}, 90 (1982).

\bibitem{ab} 
M. Abramowitz and I. A. Stegun (eds.),
{\it Handbook of Mathematical Functions\/} 
(Dover: New York 1970).

\bibitem{la59} 
L. D. Landau and E. M. Lifshitz, 
{\it Statistical Physics\/}, p. 162f 
(Pergamon Press: London 1959).

\bibitem{ca68} 
V. Canuto and H.-Y. Chiu, Phys. Rev. {\bf 173}, 1220 (1968).
 
\bibitem{ka91} 
A. D. Kaminker, K. P. Levenfish, and D. G. Yakovlev,
Sov. Astron. Lett. {\bf 17}, 450 (1991). 

\bibitem{fr79}
B. L. Friman and O. V. Maxwell,
Astrophys. J. {\bf 232}, 541 (1979).

\bibitem{sh83}
S. L. Shapiro and S. A. Teukolsky,
{\it Black Holes, White Dwarfs, and Neutron Stars\/}
(Wiley: New York 1983).

\bibitem{st91}  
N. Straumann, 
{\it General Relativity and Relativistic Astrophysics\/}
(Springer-Verlag: Berlin 1991).

\bibitem{ne73} 
J. W. Negele and D. Vautherin, 
Nucl. Phys. {\bf A207}, 298 (1973).

\bibitem{pe94}
C. J. Pethick and V. Thorsson, Phys. Rev. Lett.  {\bf 72}, 1964 (1994).

\bibitem{be95} 
W. Becker, Ann. N.Y. Acad. Sci. {\bf 759}, 250 (1995).

\bibitem{th93}
C. Thompson and R. C. Duncan, Astrophys. J. {\bf 408}, 194 (1993).

\bibitem{sa89}
For a review see e.g. J. A. Sauls in: 
H. \"{O}gelman and E. P. J. van den Heuvel (eds),
{\it Timing Neutron Stars\/},  p. 457 
(Kluwer Academic Publishers, Dordrecht 1989).

\bibitem{ka97}
A. D. Kaminker,  D. G. Yakovlev and P. Haensel,
preprint astro-ph/9702155, to appear in Astron. Astrophys. 1997.

\bibitem{me52}
L. Mestel, Mon. Not. R. Astron. Soc.  {\bf 112}, 583 (1952).

\bibitem{sc95}
G. D. Schmidt and P. S. Smith, Astrophys. J. {\bf 448}, 305 (1995).


\end{references}
\end{document}